\documentclass[aps,prx,reprint,superscriptaddress,longbibliography,floatfix]{revtex4-2}

\newcommand{\bra}[1]{\langle #1 |}
\newcommand{\ket}[1]{| #1 \rangle }

\usepackage{amsmath}
\usepackage{amstext}
\usepackage{amssymb}
\usepackage{graphicx}
\usepackage{import}
\usepackage{appendix}
\usepackage{lipsum}

\usepackage[pdfpagemode=UseNone,pdfstartview=FitH,colorlinks=true,linkcolor=blue,citecolor=blue,urlcolor=blue]{hyperref}
\usepackage[all]{hypcap}
\usepackage{subfigure}

\begin{document}

\title{Measurement-Induced State Transitions in a Superconducting Qubit: Within the Rotating Wave Approximation}

\newcommand{\Google}{Google Quantum AI, Goleta, CA}
\newcommand{\UCR}{Department of Electrical and Computer Engineering, University of California, Riverside, CA}
\newcommand{\UCSB}{Department of Physics, University of California, Santa Barbara, CA, USA}

\author{Mostafa Khezri}
\thanks{Equal contribution}
\author{Alex Opremcak}
\thanks{Equal contribution}
\author{Zijun Chen}
\author{Kevin C. Miao}
\affiliation{\Google}
\author{Matt McEwen}
\affiliation{\Google}
\affiliation{\UCSB}
\author{Andreas Bengtsson}
\author{Theodore White}
\author{Ofer Naaman}
\author{Daniel Sank}
\affiliation{\Google}
\author{Alexander N. Korotkov}
\affiliation{\Google}
\affiliation{\UCR}
\author{Yu Chen}
\author{Vadim Smelyanskiy}
\affiliation{\Google}

\begin{abstract}
Superconducting qubits typically use a dispersive readout scheme, where a resonator is coupled to a qubit such that its frequency is qubit-state dependent.
Measurement is performed by driving the resonator, where the transmitted resonator field yields information about the resonator frequency and thus the qubit state.
Ideally, we could use arbitrarily strong resonator drives to achieve a target signal-to-noise ratio in the shortest possible time.
However, experiments have shown that when the average resonator photon number exceeds a certain threshold, the qubit is excited out of its computational subspace in a process we refer to as a measurement-induced state transition (MIST).
These transitions degrade readout fidelity, and constitute leakage which precludes further operation of the qubit in, for example, error correction.
Here we study these transitions experimentally with a transmon qubit by  measuring their dependence on qubit frequency, average resonator photon number, and qubit state, in the regime where the resonator frequency is \emph{lower} than the qubit frequency.
We observe signatures of resonant transitions between levels in the coupled qubit-resonator system that exhibit noisy behavior when measured repeatedly in time.
We provide a semi-classical model of these transitions based on the rotating wave approximation and use it to predict the onset of state transitions in our experiments.
Our results suggest the transmon is excited to levels near the top of its cosine potential following a state transition, where the charge dispersion of higher transmon levels explains the observed noisy behavior of state transitions.
Moreover, we show that occupation in these higher energy levels poses a major challenge for fast qubit reset.
\end{abstract}

\maketitle

\section{Introduction}\label{sec:intro}
Superconducting qubits typically use the dispersive interaction with a linear resonator for state measurement \cite{Blais2004, Wallraff2004, Blais2021}.
This interaction is modeled by the Jaynes-Tavis-Cummings (JTC) Hamiltonian \cite{Jaynes1963, Tavis1968} that can generally describe the coupling between a multi-level nonlinear system (qubit) and a linear resonator, where the frequency of the resonator depends on the state of the nonlinear system.
The resonator frequency is probed to infer the state of the qubit.
Dispersive readout has been the method of choice for most superconducting qubit hardware due to its speed, accuracy, and quantum non-demolition character \cite{Johnson2012, Walter2017, Arute2019, Jurcevic2021}.
These properties are particularly important for cyclic error correcting protocols that require the qubit state to remain in a known computational state after measurement and reset are complete, and where the cycle time needs to be much shorter than the characteristic coherence time of the qubits \cite{Fowler2012, Chen2021, Krinner2022, Acharya2022}.

A straightforward approach for making dispersive readout faster is to increase the strength of the resonator drive pulse, which increases the average (resonator) photon number.
However, multiple experiments have found that when the average photon number exceeds a certain threshold, the qubit undergoes a state transition that degrades readout fidelity and leaves the qubit in a state outside of the computational subspace \cite{Reed2010, Sank2014, Lescanne2019}, thus limiting measurement speed.
Ref.~\cite{Khezri2016} found that, when the resonator frequency is \emph{higher} than the qubit frequency, state transitions are \emph{not} predicted by JTC interactions within the rotating wave approximation (RWA).
Ref.~\cite{Sank2016} then showed that in this regime the transitions are mediated by non-RWA interactions instead, and are caused by resonances within the overall structure of the JTC energy ladder.

Dispersive readout can also be performed when the resonator frequency is \emph{lower} than the qubit frequency.
This regime is of practical significance since a frequency-tunable qubit can be brought into resonance with the resonator to remove leakage outside of the computational subspace and reset the qubit to ground state \cite{McEwen2021}.
Such leakage removal reset schemes are typically easier to calibrate and also faster than other methods where the qubit and resonator are detuned, e.g., in the case where resonator frequency is higher than the qubit's maximum frequency \cite{Geerlings2013, Magnard2018, Zhou2021, Marques2023}.
Furthermore, evidence shows that fast and accurate removal of leakage is a necessary component for achieving performant error correcting codes \cite{Miao2022}.

In this work we study measurement-induced state transitions (MIST) using a transmon qubit \cite{Koch2007} when the resonator frequency is \emph{lower} than the qubit frequency.
We begin by characterizing the dependence of state transitions on qubit frequency, average photon number, and qubit state.
We then provide a semi-classical model that predicts the onset of transitions versus qubit frequency and photon number with no fitting parameters.
Similar to Ref. \cite{Sank2016}, we find that the state transitions are caused by energy resonances within the JTC energy ladder, but in this regime they are mainly mediated by RWA interaction terms and can occur at relatively small photon numbers $\sim 5$.
Additionally, we find that transmon offset charge changes these resonance conditions, thus explaining the fluctuations in these transitions when measured repeatedly in time.
Finally, we show that when a qubit undergoes a measurement-induced state transition, the leaked state is resistant to reset and therefore poses complications for further operations of the system, specifically cyclic error correcting codes.

\section{Experiment}\label{sec:exp}
\begin{figure}[]
\centering
\includegraphics[scale=0.75]{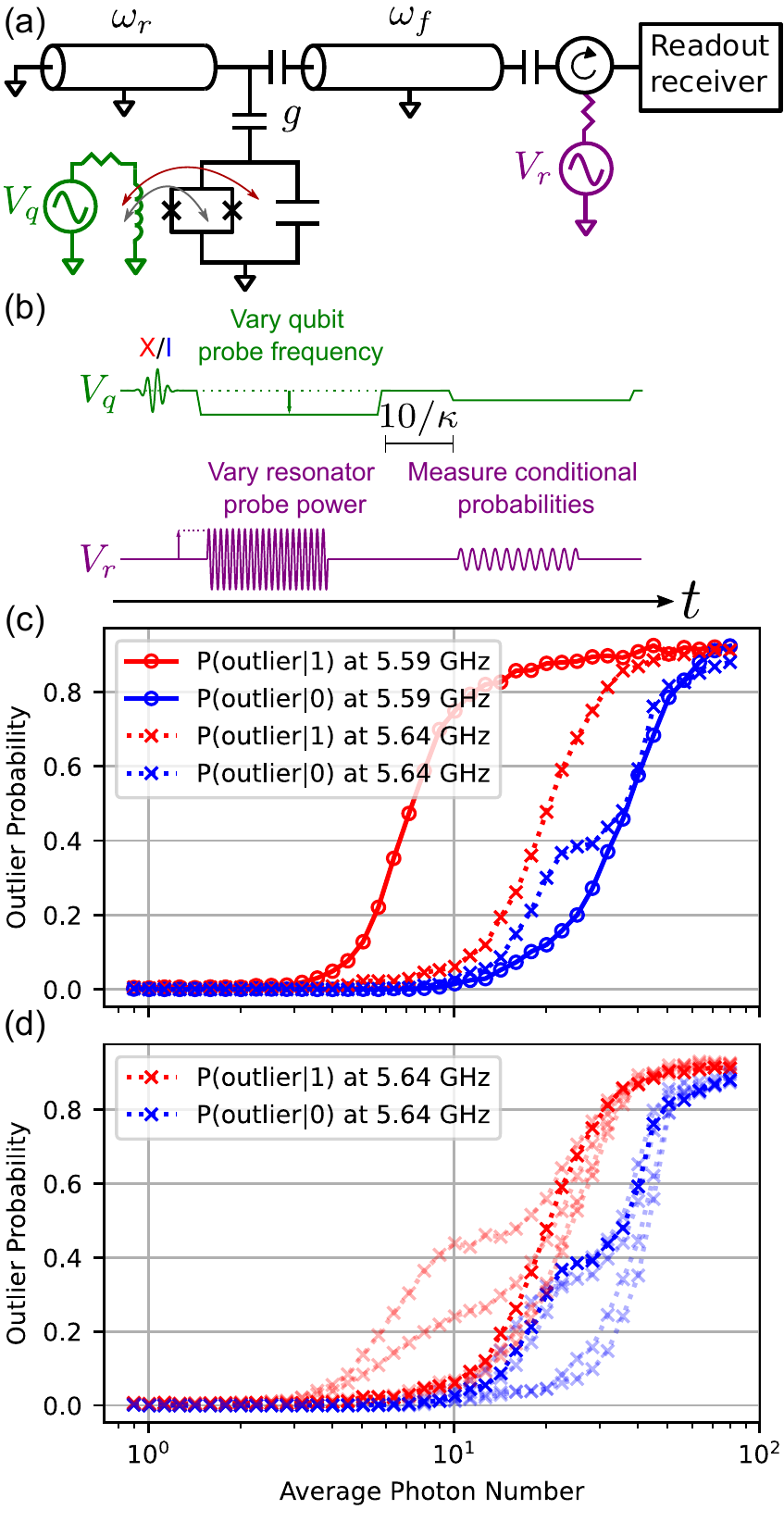}
    \caption{Experimental setup and initial observations. \textbf{(a)} Circuit schematic of the qubit-resonator system.
    \textbf{(b)} Pulse sequence used for characterizing state transitions. The x- and y-axis are not drawn to scale; see text for timing information.
    \textbf{(c)} Conditional outlier probability versus average photon number with the qubit prepared in $\ket{0}$ (blue curves) or $\ket{1}$ (red curves) at two different qubit probe frequencies (solid vs dashed). \textbf{(d)} With the qubit probe frequency fixed at 5.64 GHz, we repeat the experiment from panel (c) in time. Each semi-transparent curve corresponds to a unique realization of the experiment. The opaque curves correspond to the data from panel (c) with the qubit probe frequency at 5.64 GHz.
    Error bars are smaller than the marker size.}
    \label{fig:1}
\end{figure}

Our circuit consists of a frequency-tunable transmon qubit that is dispersively coupled to a Purcell-filtered resonator \cite{Reed2010_purcell, Jeffrey2014, Sank2014}, with bare resonator frequency $\omega_r/(2\pi)=4.750 ~\text{GHz}$, Purcell filter frequency $\omega_f \approx \omega_r$, qubit-resonator coupling strength $g/(2\pi)\approx 130~\text{MHz}$, and resonator decay rate $\kappa = 1/(22~\text{ns})$ as shown in Fig. \ref{fig:1}(a).
The qubit has a maximum frequency $\omega_{q,\text{max}}/(2\pi) = 6.340~\text{GHz}$, nonlinearity $\eta \equiv \omega_q - \omega_{21} \approx 2\pi \times 195 ~\text{MHz}$, and is operated at frequencies above the resonator in these experiments ($\Delta = \omega_q - \omega_r > 0$), in contrast to Ref.~\cite{Sank2016} where $\Delta < 0$.
The qubit and resonator are controlled using external voltage sources $V_q$ and $V_r$, respectively, which are responsible for single-qubit gates (XY), qubit frequency tuning (Z), and resonator drive pulses.
Readout signals scattered off of the resonator-filter system are amplified, digitized, and post-processed by the receiver, providing measurement of qubit states $\ket{0}$ and $\ket{1}$.
In what follows, we refer to any state outside of the 01-subspace as an \textit{outlier}.

We characterize state transitions using the pulse sequence depicted in Fig. \ref{fig:1}(b).
At the beginning of each measurement repetition, the qubit is tuned to its resonator frequency for 1~$\mu\text{s}$ to prepare the qubit in state $|0\rangle$ (waveform not shown) \cite{Reed2010_purcell}.
Next, the qubit is moved to its idle frequency and an $X$-gate ($I$-gate) is used to prepare the qubit in state $\ket{1}$ ($\ket{0}$).
The qubit is then tuned to its \textit{probe} frequency, at which point the resonator is driven using a probe tone that is resonant with the corresponding dressed resonator frequency for a duration of $1~\mu\text{s}$ ($\gg 1/\kappa$) to ensure that the system reaches a steady state.
See Appendix~\ref{app:dressed_tracking} for information on how the dressed resonance frequencies were measured in these experiments.
At the end of the resonator drive, the system idles for 400 ns to allow the resonator photons to decay.
Then, the qubit is tuned back to its idle frequency, where the system idles for an additional $10/\kappa = 220~\text{ns}$ to ensure the resonator has returned to vacuum.
Finally, the qubit is measured to extract the conditional probabilities of finding the qubit in $\ket{0}$, $\ket{1}$, or an outlier state, given the initial state, i.e., $P(\text{final}|\text{initial})$.
For each combination of the initial qubit state, qubit probe frequency, and resonator probe power, the experiment is repeated 5,000 times to estimate conditional probabilities.

In Fig. \ref{fig:1}(c), we show the conditional outlier probability versus the average photon number with the qubit prepared in $\ket{0}$ (blue) or $\ket{1}$ (red).
See Appendix~\ref{app:photon_estimate} for information about conversion between resonator probe power and average photon number.
We show data with the qubit probe frequency at 5.59 GHz (solid lines, $\circ$ markers) and 5.64 GHz (dashed lines, $\times$ markers).
In Fig. \ref{fig:1}(d), we repeat the experiment from Fig. \ref{fig:1}(c) in time, with the qubit probe frequency fixed at 5.64 GHz.
The semi-transparent curves, which follow the same color, marker, and line style conventions described in Fig. \ref{fig:1}(c), correspond to three additional realizations of this experiment, taken consecutively with one minute between each realization, where each realization takes $\sim 1$ second to complete. The opaque curves correspond to the data from Fig. \ref{fig:1}(c) with the qubit probe frequency at 5.64 GHz.
We observe several interesting trends: (i) the onset of state transitions typically occurs at a lower average photon number for $\ket{1}$ than $\ket{0}$, (ii) features in outlier probability are strongly dependent on the qubit probe frequency, and (iii) features in outlier probability change significantly on timescales $\sim 1$ minute.

\section{Model}\label{sec:model}
We model our experiment using a semi-classical approach where the resonator is described by a coherent state (classical field) that directly drives the transmon via RWA interactions.
The Hamiltonian of our model is ($\hbar=1$, see Appendix~\ref{app:theory})
\begin{align}
    H = & \sum_k (E_k - k\omega_r) \ket{k}\bra{k} \label{eq:H-direct_drive}\\
     + & \frac{\alpha e^{i(\omega_r - \omega_d)t}}{|\alpha|} \sum_k \text{Re}\left( \sqrt{|\alpha|^2 - k}\right)g_{k, k+1}\ket{k+1}\bra{k} \nonumber \\
     + & \text{H.c.} \nonumber
\end{align}
where $E_k$ ($\ket{k}$) are the transmon eigenenergies (eigenstates) and $g_{k, k+1} \equiv g \bra{k} \hat{n} \ket{k+1} / \bra{0} \hat{n} \ket{1}$ are the normalized transmon charge matrix elements in its eigenbasis, where $g = k_\text{eff}\sqrt{\omega_q \omega_r}/2$ is the qubit-resonator coupling strength defined via coupling efficiency $k_\text{eff}$.
We note that $E_k$, $\ket{k}$, and $g_{k, k+1}$ depend on offset charge.
Lastly, $\alpha$ describes the coherent state of the resonator and evolves as
\begin{equation}\label{eq:coherent}
    \dot{\alpha}(t) = -i(\tilde{\omega}_r - \omega_d)\alpha(t) - \frac{\kappa}{2} \,\alpha(t) -i\varepsilon,
\end{equation}
where $\kappa$ is the resonator's energy decay rate, $\varepsilon$ and $\omega_d$ denote the amplitude and frequency of the resonator drive, respectively, and $\tilde{\omega}_r$ is the dressed resonator frequency corresponding to the initial qubit state.
Note that the average photon number in the coherent state is $\bar{n}=|\alpha(t)|^2$ which fluctuates by $\pm \sqrt{\bar{n}}$, and this uncertainty is always present in our model when predicting the onset of readout transitions.
The term $\text{Re}\left( \sqrt{|\alpha|^2 - k}\right)$ is used to turn off the interaction when $\bar{n}<k$, which mimics the behavior of the full JTC model at small photon numbers. See Appendix~\ref{app:theory} for details of the model.

\begin{figure}[t]
\begin{center}
    \includegraphics[width=1\columnwidth]{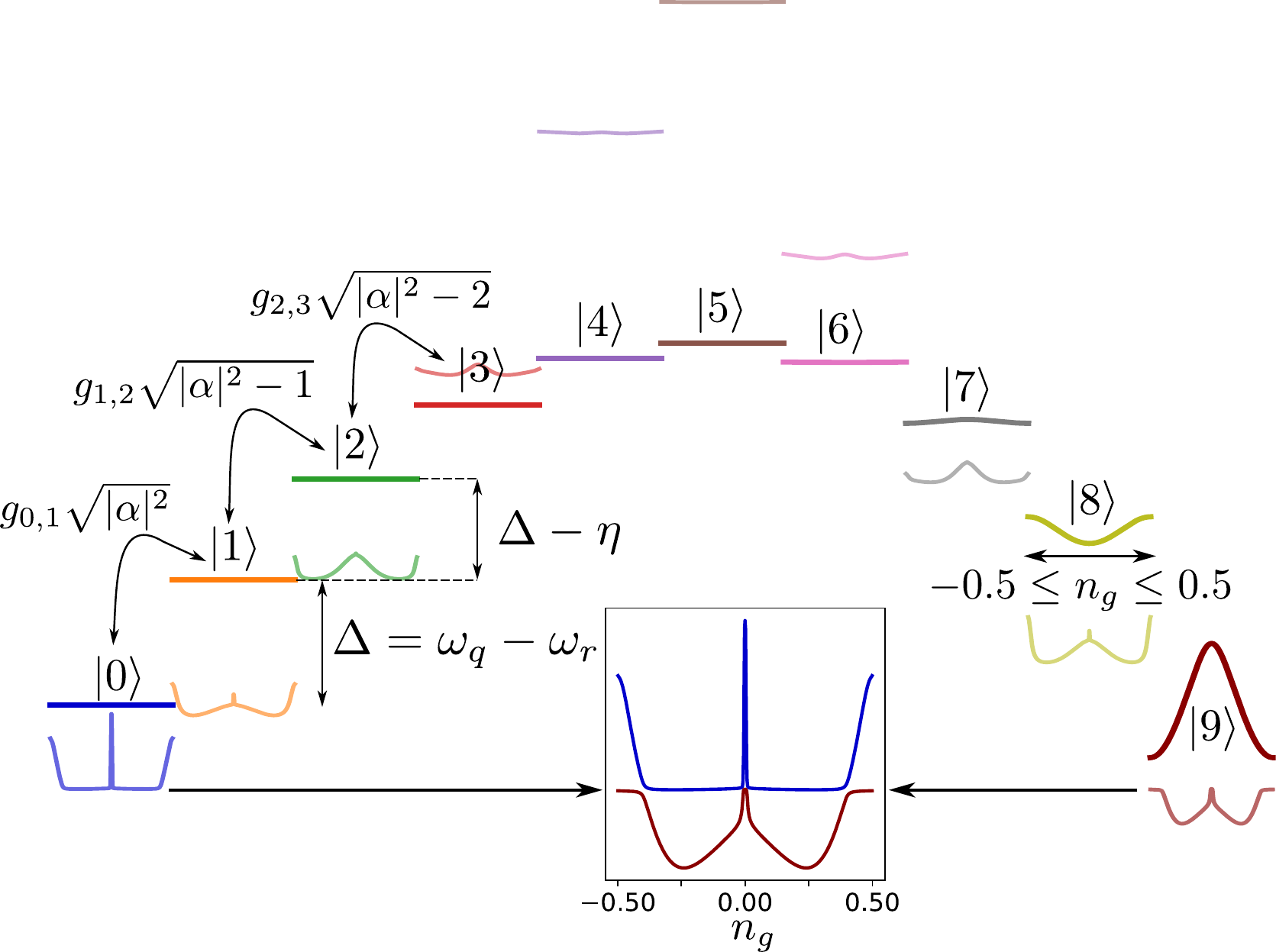}
\end{center}
    \caption{Energy structure of an RWA strip with $\omega_q > \omega_r$ for $\bar{n}=|\alpha|^2 = 50$.
    Thick lines show the bare energies, thin lines show the instantaneous eigenenergies, and the offset charge is scanned along the length of each line in the range corresponding to one period $-0.5\le n_g \le 0.5$.
    Note that eigenenergies at the bottom of the well have a noticeable offset charge dispersion compared to their bare energies.
    Curved arrows show the coupling between bare levels, and straight vertical arrows show the energy difference between them.
    The panel shows resonance (i.e., avoided crossing) between $\ket{0}$ and $\ket{9}$ at multiple offset charge values.
    Not all levels are annotated or shown.
    Here we use $\Delta/2\pi = 1$ GHz, $\eta/2\pi = 200$ MHz, and $g/2\pi = 120$ MHz.
    }
    \label{fig:2}
\end{figure}

We use our model to simulate the behavior of the qubit during readout as follows.
First, we solve Eq.~\eqref{eq:coherent} for the coherent state evolution starting with $\alpha(t=0)=0$, where we use a resonant drive $\omega_d = \omega_r$ to replicate the experimental procedure (see Appendix~\ref{app:dressed_tracking}).
We then plug this solution into Eq.~\eqref{eq:H-direct_drive} and numerically solve the Schr\"{o}dinger equation for the state of the transmon, starting from initial states $\ket{0}$ or $\ket{1}$.
Here, the resonator field plays the role of a classical drive that ramps up the interactions between levels as photon number increases, where solving the Schr\"{o}dinger equation for $\sim 20$ transmon levels is fast.
Finally, we look at the populations of the instantaneous eigenstates of the system; a state transition in our model is identified by loss of the initial population, and that is what we will compare later with experiments.
See Appendix~\ref{app:theory} for an example of such population loss.

Fig.~\ref{fig:2} shows an example of the energy structure of the model in Eq.~\eqref{eq:H-direct_drive}.
Our model effectively describes the evolution within an RWA strip \cite{Sank2016}, i.e., levels that are coupled via excitation preserving interaction terms of the JTC ladder.
Note that when $\omega_r < \omega_q$, the RWA strip bends over itself at $k_\text{bend} \approx (\omega_q - \omega_r)/\eta$ and levels in the bottom of the transmon cosine potential can become on resonance with levels near the edge of the cosine potential, i.e., the qubit may borrow a few excitations from the resonator and jump upward several levels.
The coupling between these resonant levels is provided by a multi-step process involving intervening virtual levels.
This is the main mechanism that leads to state transitions and transfers the qubit population from the computational subspace to higher states.

Note that the large offset charge dispersion of transmon levels near the edge of the cosine potential leads to a significant role of offset charge for the resonance conditions studied here.
Additionally, we observe that the offset charge dispersion of these higher levels affects not only their own eigenenergies but also the eigenenergies of levels near the \emph{bottom} of the well.
This is depicted in Fig.~\ref{fig:2}, where thick lines show the bare energies of the system and thin lines show the instantaneous eigenenergies, and the offset charge is scanned along the length of each line over a full period of $n_g \in [-0.5, 0.5]$ to show the dispersion of each level.

In the full JTC energy ladder, different RWA strips are detuned by integer multiples of the resonator frequency and are coupled to each other via non-RWA terms.
Due to the energy structure of the system in this parameter regime, in contrast to the case of Ref.~\cite{Sank2016}, it is less likely for different RWA strips to become on resonance with each other at small photon numbers.
We have checked that with the addition of non-RWA terms the eigenenergies and the location of resonances change but not significantly.
Therefore, in this work we focus on dynamics within RWA and defer study of non-RWA effects to a future work.

Our main goal with this model is computational speed and applicability: we need to predict the occurrence of state transitions versus qubit frequency and photon number to avoid them during readout, and to inform parameters for new designs.
As such, the model presented in this paper was used to optimize the readout in Refs.~\cite{Acharya2022, Bengtsson2023}.
However, this is an approximate model that is no longer applicable after a transition occurs, as the resonator is no longer described by a coherent state.
Ref.~\cite{Shillito2022} performed fully quantum simulations of the system with a very large Hilbert space, which can be used to study the dynamics during and after a state transition occurs.
They also provided a semiclassical method for studying these systems based on evolution of the resonator that is dressed by the qubit.
Recently, Ref.~\cite{Cohen2022} studied the same problem using a directly driven transmon model, where they used the alternative approach of Floquet solutions, with an emphasis on identifying signatures of chaos in the transmon.
They found results similar to our work such as resonance between transmon levels due to the drive, increased offset charge dispersion of lower levels in presence of the drive, and the important role of offset charge in calculating system parameters.

\section{Experiment vs model}\label{sec:exp_vs_model}

\begin{figure}[t]
\begin{center}
    \includegraphics[width=\columnwidth]{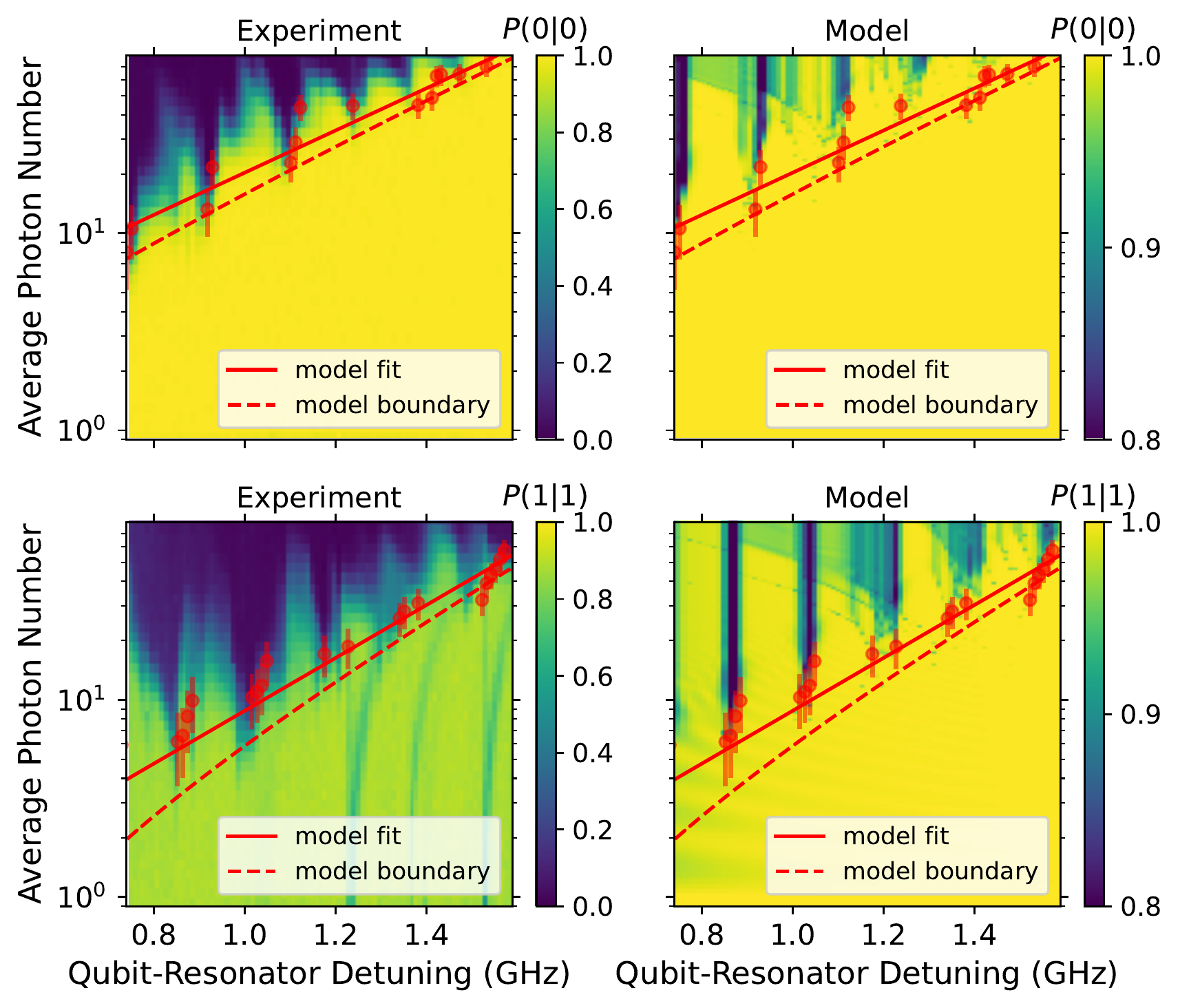}
\end{center}
    \caption{Probability of remaining in the initial state for ground (top row) and excited (bottom row) state as a function of qubit-resonator detuning and resonator average photon number, as observed in experiment (left column) and the model (right column).
    Red filled circles mark the photon numbers at which state transitions occur in the model, with error bars $\pm\sqrt{\bar{n}}$, and they are re-drawn on the experimental data for easier comparison.
    Solid red lines show an exponential fit to the filled circles as $n_\text{fit}=A e^{B\Delta}$, and the dashed red lines show the boundary of state transitions defined as $n_\text{fit} - \sqrt{n_\text{fit}}$.
    The model uses transmon charging energy of $194$ MHz, $k_\text{eff}=0.048$, $1/\kappa=22$ ns, $\omega_r/2\pi=4.750$ GHz, $\varepsilon/2\pi=45$ MHz, $\omega_d=\omega_r$, and the coherent state is evolved for 100 ns under the drive.
    In the model we truncate the transmon at 20 eigenstates, tune transmon junction energy to sweep the qubit frequency, and take a uniform average over the offset charge in the range $n_g \in [-0.5, 0]$.
    In the experimental data for $\text{P}(1|1)$, the vertical features extending down to the lowest photon numbers in the 1.2-1.6 GHz range are due to resonant interactions with two level system defects.
    }
    \label{fig:3}
\end{figure}

We capture the trends in state transitions by repeating our experiment over a range of qubit frequencies for initial states $\ket{0}$ and $\ket{1}$.
The result is presented in the left column of Fig.~\ref{fig:3}, where the heatmap shows the probability of remaining in the prepared state.
To compare with the experimental data, we simulate the model over the same range of qubit frequencies and initial states, using experimentally measured parameters of the qubit-resonator system.
To capture the effect of the offset charge, the model heatmaps are uniformly averaged over simulations for $n_g \in [-0.5, 0]$ (the behavior is symmetric around $n_g=0$) in steps of $0.05$.
The model heatmaps are shown in the right column of Fig.~\ref{fig:3}. 
Note that the experiment is sampling the offset charge in a manner that depends on its noisy behavior \cite{Riste2013, Christensen2019}, and therefore repeating the same experiment in time can yield different results from run to run, in agreement with the observations in Fig.~\ref{fig:1}(d).

Comparing the model with the experimental data in Fig.~\ref{fig:3} shows good qualitative and quantitative agreement both in detuning and photon number at which state transitions occur, where features group together in frequency near bands of resonance.
Roughly speaking, each resonance band corresponds to a detuning range which bends the RWA strip at a specific level.
This is because the eigenenergies within an RWA strip depend strongly on the level at which the strip bends over itself, which is determined by the qubit-resonator detuning.
As such, it is evident that the multi-level energy structure of the qubit circuit plays an important role in resonance conditions, which depends on the details of the transmon circuit model \cite{Khezri2018}, estimates of the circuit parameters such as charging energy, and also dressing by other nearby elements (e.g., couplers).
Note that some of the experimentally observed transitions in between the resonance bands and at higher photon numbers, which are absent in the model, can originate from non-RWA interactions ignored in this model; however, as evident from Fig.~\ref{fig:3}, they are not the main effect that limits the allowed photon number for the readout in this regime.
We also compared the model with experiments performed on different qubit-resonator systems that were fabricated with coupling efficiencies in the range 0.025--0.04 and anharmonicities in the range 190--280 MHz, and we observed similar agreement.

As we show in Sec.~\ref{sec:reset}, measurement-induced transition to higher states is mostly immune to typical reset protocols \cite{McEwen2021} and leaves the qubit excited for a long time, which is detrimental to cyclic operations of quantum circuits.
Therefore it is of practical interest to predict the photon number at which transitions set in.
To do that, in our simulations we identify the photon number for the onset of state transitions where the initial probability drops below a given threshold, and only keep the points that monotonically increase with detuning.
These points are marked with red filled circles in Fig.~\ref{fig:3} with error bars indicating $\pm\sqrt{\bar{n}}$, and are re-drawn on the experimental data for easier comparison.
By fitting a phenomenological exponential model $n_\text{fit}=A e^{B\Delta}$ to these points, (solid red lines in Fig.~\ref{fig:3}), we define the boundary of state transitions as $n_\text{fit} - \sqrt{n_\text{fit}}$ (dashed red lines in Fig.~\ref{fig:3}) and avoid operating beyond this boundary.
The boundary depends on the qubit-resonator detuning $\Delta$, coupling between them $g$, and the qubit anharmonicity $\eta$, as these are the main parameters that shape the eigenenergies of an RWA strip.
Note that for our parameters, the photon number at which transitions set in is $\sim3\times$ lower for excited state than for ground state, and that sets the photon limit for dispersive measurement of unknown states.

\section{Qubit reset following a state transition}\label{sec:reset}
Our theory suggests that state transitions may leave the transmon in energy levels near the top of the cosine potential at the end of the measurement process. 
These higher transmon energy levels are difficult to reset as: (i) their transition frequencies are extremely sensitive to offset charge (with charge dispersions $\sim$100 MHz), (ii) their transition frequencies can be significantly lower than $\omega_{q}$ and $\omega_r$, and (iii) the number of excitations needed to be removed from the qubit to achieve reset is large ($\sim$ 5-10).
These considerations pose major complications for microwave- and resonator-assisted qubit reset schemes \cite{Reed2010_purcell, Geerlings2013, Magnard2018, Zhou2021}.

\begin{figure}[ht!]
\centering
\includegraphics[scale=0.9]{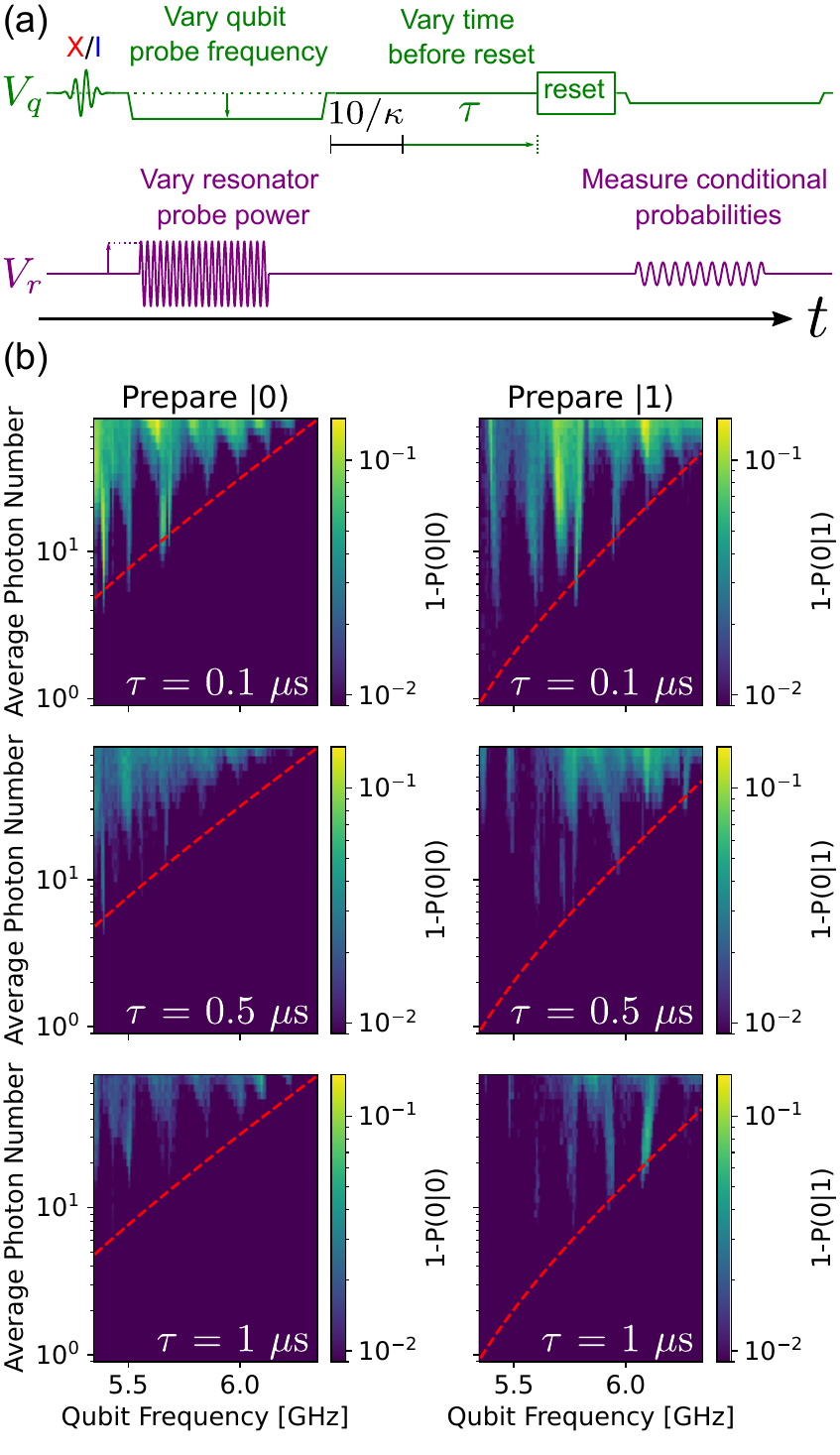}
    \caption{Qubit reset following a state transition. \textbf{(a)} Pulse sequence used for characterizing qubit reset error following a measurement-induced state transition. The x- and y-axis are not drawn to scale; see text for timing information. \textbf{(b)} Reset error versus qubit probe frequency and average photon number. The left (right) column of images corresponds to the prepared state $\ket{0}$ ($\ket{1}$), with a time before reset $\tau$ = 0.1 $\mu$s (top), 0.5 $\mu$s (middle), and 1 $\mu$s (bottom). The dashed red line in each image is the state transition boundary corresponding to the prepared qubit state. }
    \label{fig:4}
\end{figure}

We characterize our ability to reset the qubit following a state transition using the multi-level reset (MLR) gate developed in Ref. \cite{McEwen2021}.
In the absence of a resonator probe pulse, we measure a \textit{reset error}, defined as the probability of not finding the qubit in $\ket{0}$, of $<0.5\%$ when the qubit is initially prepared in state $\ket{0}$, $\ket{1}$, or $\ket{2}$. 
To characterize reset error following a state transition, we modify our original pulse sequence [Fig. \ref{fig:1}(b)] by placing an idle gate, with duration $\tau$, followed by an MLR gate before measurement of the conditional probabilities [see Fig. \ref{fig:4}(a)]. 
In Fig. \ref{fig:4}(b), we show the measured reset error (color scale) versus the average photon number and qubit probe frequency, at idle gate durations of $\tau$ = 0.1 $\mu$s (top row), $\tau$ = 0.5 $\mu$s (middle row), and $\tau$ = 1 $\mu$s (bottom row).
The left (right) column corresponds to the prepared state $\ket{0}$ ($\ket{1}$). 
We detect significant levels of reset error ($\gtrsim 4\%$) above the state transitions boundary (dashed red lines) for $\tau = 0.1,~0.5,~1~\mu\text{s}$.
For comparison of time scales, the average qubit relaxation time over this frequency range is $T_{1\rightarrow 0} \approx 20 \,\mu s$, and relaxation time for higher transmon levels is roughly $T_{k\rightarrow k-1} \sim T_{1\rightarrow 0}/k$ \cite{Peterer2015}.
This demonstrates that state transitions produce long-lived leakage states that are immune to the MLR gate, indicating significant qubit population above state $\ket{2}$.
Such leakage states are detrimental to error correcting protocols \cite{Miao2022} and were intentionally avoided in Ref. \cite{Acharya2022} by choosing readout parameters that fall below state transitions boundary.

\section{Conclusion}\label{sec:conclusion}
In conclusion, we have characterized measurement-induced state transitions in a transmon qubit when the readout resonator frequency is lower than the qubit frequency.
With $\omega_r < \omega_q$, we find that state transitions are caused by resonances in the JTC energy ladder that are mediated by RWA interactions: the transmon's negative anharmonicity enables the qubit to borrow a few excitations from the resonator and transition to a higher state in an excitation preserving manner.
These findings are unlike the ones in Ref.~\cite{Sank2016}, which studied a transmon-resonator system with similar parameters, but with $\omega_q < \omega_r$, and found that non-RWA interactions within the JTC Hamiltonian were needed to explain state transitions.

We provide a semi-classical model that predicts the onset of state transitions versus qubit frequency, average photon number, and qubit state with no fitting parameters.
We also observe that due to large offset charge dispersion of the higher transmon levels, the role of offset charge becomes important in explaining experimental observations in two ways: (i) it explains why experimental features vary in time, and (ii) averaging simulations over offset charge is required to properly compare them with experiments.
Our results show that the onset of state transitions is exponentially suppressed with increased detuning, though the resonator dispersive shift becomes quadratically smaller.
We also establish that the excited state is more susceptible to state transitions versus average photon number, and that sets the photon limit for dispersive measurement.
We also note that following a state transition event, fast reset protocols fail to reset the qubit, and that can be detrimental to error correcting algorithms.

\begin{acknowledgments}
We are grateful to the Google Quantum AI team for building, operating, and maintaining software and hardware infrastructure used in this work.
\end{acknowledgments}

\appendix

\section{Dressed resonance tracking}\label{app:dressed_tracking}
\begin{figure}[ht!]
\centering
\includegraphics[scale=0.9]{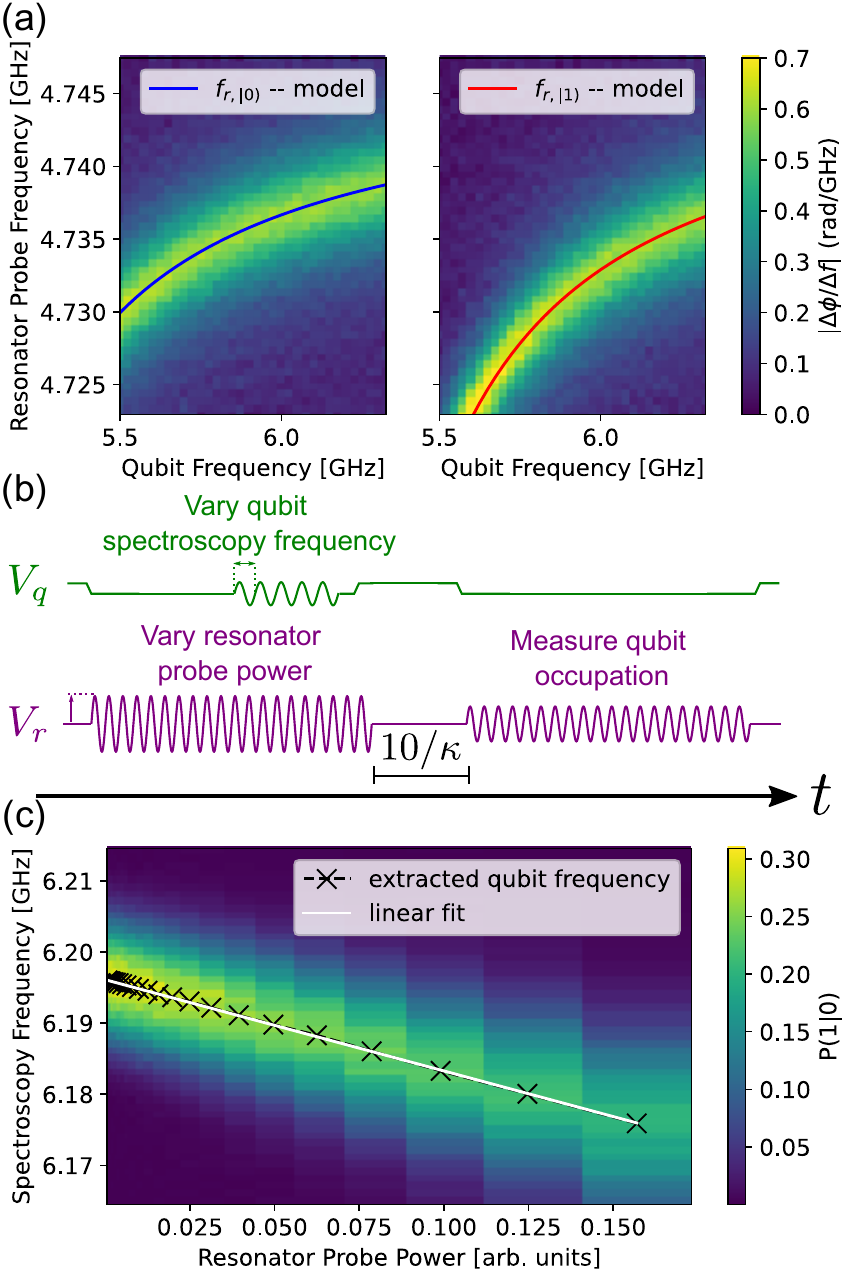}
    \caption{Resonator frequency tracking and photon number estimates. 
    \textbf{(a)} Resonator spectroscopy versus qubit frequency with the qubit prepared in $\ket{0}$ (left panel) and $\ket{1}$ (right panel). 
    The blue (red) solid line is a model of the dressed resonator frequency $\omega_{r,\ket{0}}$ ($\omega_{r,\ket{1}}$) as a function of qubit frequency. 
    \textbf{(b)} Control and measurement sequence used for $\bar{n}$ calibration. The x- and y-axis are not drawn to scale; see text for timing information.
    \textbf{(c)} AC Stark spectroscopy versus resonator probe power. 
    The $\times$-shaped markers are the extracted qubit (center) frequencies versus resonator probe power. 
    The data are well described by a linear model (see Eq. (\ref{eq:linear_ac_stark})).
    \label{fig:S1}}
\end{figure}

To keep the range of average photon number, $\bar{n}$, constant as the qubit probe frequency is varied in our experiments, we must account for changes in the dressed resonance frequencies, $\omega_{r,\ket{0}}$ and $\omega_{r,\ket{1}}$. We characterize this with resonator spectroscopy versus qubit frequency as shown in Fig. \ref{fig:S1}(a). 
In this experiment, the qubit is prepared in $\ket{0}$ or $\ket{1}$, and then rapidly tuned (with risetime $\sim$ 1-2 ns) to the frequency of interest.
Next, a variable frequency resonator spectroscopy pulse is applied to the resonator-filter system (see Fig. \ref{fig:1}(a) for circuit information).
Note that this spectroscopy pulse uses a small power to prevent the resonator from exhibiting nonlinear behavior due to its coupling to the qubit.
The scattered signal is then heterodyne-detected by the readout receiver, allowing for extraction of the amplitude and phase ($\phi$) versus the spectroscopy frequency ($f$). 
On resonance, the rate of change $\phi$ with respect to $f$, denoted by $\Delta \phi / \Delta f$, is maximized.
The color scale in Fig. \ref{fig:S1}(a) is $|\Delta \phi / \Delta f|$, and is evaluated along each fixed vertical line cut (i.e., at a fixed qubit frequency).
To extract models for the dressed resonator frequencies, we fit the locations of maxima in $|\Delta \phi / \Delta f|$ versus qubit frequency to the dispersive model for a transmon-resonator system \cite{Khezri2018}:
\begin{eqnarray}
    \omega_{r,\ket{0}} &=& \omega_r - \frac{g^2}{\Delta} \label{eq:dressed_0_resonance} \\
    \omega_{r,\ket{1}} &=& \omega_{r,\ket{0}} - 2 \chi,  \label{eq:dressed_1_resonance}
\end{eqnarray}
where
\begin{eqnarray}
    \chi = \frac{g^2}{\Delta} \times \bigg( \frac{\eta}{\Delta - \eta}\bigg) \times \bigg(\frac{\omega_r}{\omega_q}\bigg). \label{eq:chi}
\end{eqnarray}
The blue (red) solid line in Fig. \ref{fig:S1}(a) is a fit to Eq. (\ref{eq:dressed_0_resonance}) for $\omega_{r,\ket{0}}$ (Eq. (\ref{eq:dressed_1_resonance}) for $\omega_{r,\ket{1}}$). 
Finally, to maintain a resonant probe tone versus qubit probe frequency during our experiments, we drive the resonator at the extracted value for $\omega_{r,\ket{0}}$ ($\omega_{r,\ket{1}}$) when the qubit is prepared in $\ket{0}$ ($\ket{1}$).

\section{Estimating average photon number}\label{app:photon_estimate}
To compare our experimental results with the model, we need to present our data versus average photon number, $\bar{n}$, rather than resonator probe power, $P_\text{probe}$. 
To calibrate the conversion between $P_\text{probe}$ and $\bar{n}$, we use the ac Stark effect \cite{Schuster2005, Schuster2007}. 
The waveforms for this experiment are depicted in Fig. \ref{fig:S1}(b). 
At the beginning of each experiment, the qubit is prepared in $\ket{0}$ (waveform not shown). 
Then, the qubit is biased to the frequency of interest, at which point the resonator is driven at frequency $\omega_{r, \ket{0}}$ (see previous section for further detail).
After the steady state is reached (at $t = 500~\text{ns} \gg 1/\kappa$), a 100 ns long, variable-frequency, qubit spectroscopy pulse is applied to determine the ac-Stark-shifted qubit frequency, $\omega_q(\bar{n})$. 
At the end of the spectroscopy pulse, the resonator idles for $10/\kappa$ to allow the resonator return to vacuum. 
This is followed by a terminal measurement of the qubit $\ket{1}$ state population, $P(1|0)$. For sufficiently low $\bar{n}$, the qubit center frequency changes linearly with photon number as
\begin{equation}
    \omega_q(\bar{n}) = \omega_q(\bar{n}=0) - 2\chi \bar{n}, \label{eq:linear_ac_stark}
\end{equation}
where $\omega_q(\bar{n}=0)$ is the qubit frequency in absence of an applied drive and $\chi = (\omega_{r,\ket{0}} - \omega_{r,\ket{1}})/2 > 0$ .
In Fig. \ref{fig:S1}(c), we show the results from this experiment versus resonator probe power. 
The $\times$-shaped markers are the extracted qubit (center) frequencies at each resonator probe power, where $P(1|0)$ is the color scale. 
The extracted qubit frequencies are linear in the power (see solid white line), thus providing us with a conversion between $P_\text{probe}$ and $\bar{n}$ by dividing the measured qubit frequency shift by $2\chi$.
To avoid repeating this experiment at each qubit probe frequency, we assume the following are true: (i) $P_\text{probe}$ is constant (relative to power reading of the instruments) over the resonator frequency ranges used, and (ii) the resonator decay rate $\kappa$ is independent of the qubit frequency and state. 
Under these assumptions, the $\bar{n}$ calibration can be readily extended to the qubit $\ket{1}$ state by driving the resonator at frequency $\omega_{r, \ket{1}}$.

\onecolumngrid\
\begin{center}
\begin{figure}[h]
\includegraphics[scale=0.8]{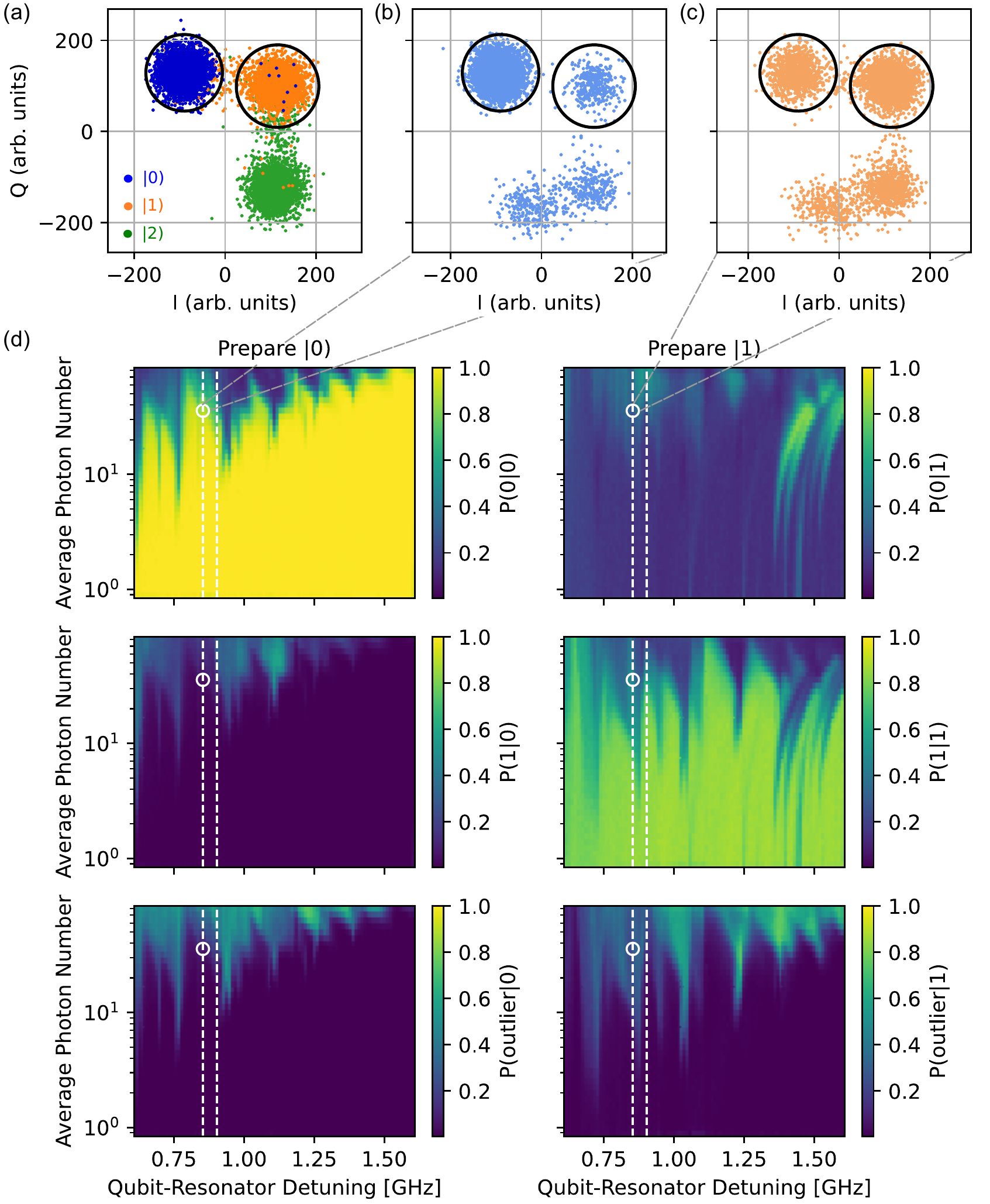}
    \caption{Characterizing state transitions versus qubit-resonator detuning and average photon number.
    \textbf{(a)} In-phase (I) and quadrature (Q) component distributions with the qubit initially prepared in state $\ket{0}$, $\ket{1}$, and $\ket{2}$.
    \textbf{(b)} IQ shots following the state transitions experiment described in Fig. \ref{fig:1} with the qubit initially prepared in $\ket{0}$.
    \textbf{(c)} Same as panel (b), but with the qubit initially prepared in $\ket{1}$. The IQ distributions in panels (b) \& (c) were taken at a qubit probe frequency of 5.59 GHz and an average photon number $\bar{n}\approx35$ photons.
    \textbf{(d)} Conditional probabilities versus qubit-resonator detuning and average photon number.
    Vertical dashed lines correspond to probed frequencies of Fig.\ref{fig:1}(c)-(d). In the panels corresponding to $\text{P}(0|1)$ and $\text{P}(1|1)$, the vertical features extending down to the lowest photon numbers in the 1.2-1.6 GHz range are due to resonant interactions with two level system defects.
    See text for further detail. 
    \label{fig:S2}}
\end{figure}
\end{center}

\twocolumngrid\

\section{Characterizing state transitions}\label{app:trans_characterize}
Characterizing measurement-induced state transitions relies on our ability to detect qubit outlier states. In Fig. \ref{fig:S2}(a), we show the in-phase (I) and quadrature (Q) component distributions with the qubit prepared in states $\ket{0}$, $\ket{1}$, and $\ket{2}$. The black circular lines surrounding the IQ distributions for the $\ket{0}$ and $\ket{1}$ states correspond to three-sigma deviations from the IQ centers and are used to define the outlier boundaries -- any shots landing outside of these regions are defined as outliers. In Fig. \ref{fig:S2}(b), we show an example of the IQ distributions with the qubit prepared in $\ket{0}$ following the state transitions experiment described in Fig. \ref{fig:1}. The IQ distributions correspond to a qubit probe frequency of 5.59 GHz and an average photon number $\bar{n}\approx35$ photons. We display the same data in Fig. \ref{fig:S2}(c), but with the qubit initially prepared in $\ket{1}$.

In Fig.~\ref{fig:S2}(d), we show conditional probabilities versus qubit-resonator detuning and average photon number. The left (right) column of panels corresponds to the initial state $\ket{0}$ ($\ket{1}$). The color scale of each panel represents the conditional probability of finding the qubit in particular final state, where the top, middle, and bottom rows correspond to final states $\ket{0}$, $\ket{1}$, and outlier, respectively. The dashed vertical lines correspond to the qubit probe frequencies used in Fig. \ref{fig:1}(c). The white circles overlaid on each panel indicate the qubit-resonator detuning and average photon number used to produce the conditional IQ distributions displayed in Fig. \ref{fig:S2}(b)-(c).
Note that the data of Fig.~\ref{fig:S2}(d) and Fig.~\ref{fig:3} were taken on the same device, but at different times.
This highlights reproducibility of our experiment and also shows the possible differences due to change of offset charge and two level system defects.

\section{Theory of direct drive of transmon}\label{app:theory}
In this section we discuss the derivation of the direct drive model used in the main text.
To model our system we start from the full Hamiltonian in the lab frame ($\hbar=1$), ignoring the Purcell filter
\begin{align}
    H = & 4E_C(\hat{n}-n_g)^2 - E_J\cos(\hat{\varphi}) \nonumber \\
    & + \omega_r \hat{a}^\dagger \hat{a} + \varepsilon e^{-i\omega_d t}\hat{a}^\dagger + \varepsilon^* e^{i\omega_d t}\hat{a} \nonumber \\
    & + ig' \hat{n} (\hat{a} - \hat{a}^\dagger). \label{eq:H-lab-s}
\end{align}
The terms in the first line of Eq.~\eqref{eq:H-lab-s} denote the transmon Hamiltonian where $E_C$ is the capacitive energy, $E_J$ is the junction energy, and $n_g$ is the offset or background charge \cite{Koch2007}.
The second line includes the terms for the resonator at frequency $\omega_r$ and its drive, where $\varepsilon$ and $\omega_d$ are the resonator drive amplitude and frequency respectively, and $\hat{a}$ denotes the resonator annihilation operator.
The third line shows the charge-charge interaction between the transmon and the resonator, where the coupling strength that is usually measured in the lab is $g =k_\text{eff}\sqrt{\omega_q \omega_r}/2$, redefined here via the coupling efficiency $k_\text{eff}$, and relates to the interaction strength in this representation as $g' \approx (32E_C/E_J)^{1/4}g$.

Next, we rewrite the Hamiltonian in the joint basis of transmon eigenbasis and resonator Fock basis, i.e., $\ket{k, n}$, where $\ket{k}$ is transmon eigenstate and $\ket{n}$ is the resonator Fock state.
We also use a rotating frame that moves the transmon to the frame of the resonator and moves the resonator to the frame of its drive, i.e., $\ket{k, n} \rightarrow e^{ik\omega_r t}e^{in\omega_d t}\ket{k,n}$.
For the charge-charge interactions, we only keep the excitation preserving terms, i.e., we employ RWA and ignore fast rotating terms, yielding
\begin{align} \label{eq:H-rf-S}
    H_\text{r.f.}^\text{RWA} = & \sum_{k,n} (E_k - k\omega_r) \ket{k, n}\bra{k, n} \nonumber \\
    & + (\omega_r - \omega_d) \hat{a}^\dagger \hat{a} + \varepsilon \hat{a}^\dagger + \varepsilon^* \hat{a} \nonumber \\
    & + \hat{a} e^{i(\omega_r - \omega_d)t} \sum_{k,n} g_{k, k+1}\ket{k+1,n}\bra{k,n} + \text{H.c.},
\end{align}
where $E_k$ and $\ket{k}$ are the transmon eigenenergies and eigenstates respectively, both of them depend on offset charge, and $g_{k, k+1} \equiv g \bra{k} \hat{n} \ket{k+1} / \bra{0} \hat{n} \ket{1}$ are the normalized transmon charge matrix elements in its eigenbasis.
Note that here $a=\sum_{k,n} \sqrt{n}\ket{k,n-1}\bra{k, n}$.

We now utilize the dressed coherent state picture \cite{Sete2013,Govia2016,Khezri2016} where it was shown that the qubit and its coupled driven resonator approximately form a coherent state made of their joint eigenstates.
In this picture, the resonator is dressed and evolves according to evolution equation for a coherent state, and the qubit is in an eigenladder that is formed due to its interaction with the resonator.
Therefore, we use an approximation that replaces the resonator operators with their classical coherent state counterpart as $\hat{a} \rightarrow \alpha(t)$.
This yields a decoupled evolution equation for the coherent state of the resonator, where the resonator field directly drives and dresses the transmon:
\begin{align}
    H_\text{eff} = & \sum_k (E_k - k\omega_r) \ket{k}\bra{k} \label{eq:H-direct_drive-s} \\
    & + \alpha(t) e^{i(\omega_r - \omega_d)t} \sum_k g_{k, k+1}\ket{k+1}\bra{k} + \text{H.c.}, \nonumber \\
    \dot{\alpha}(t) = & -i(\tilde{\omega}_r - \omega_d)\alpha(t) - \frac{\kappa}{2} \,\alpha(t) -i\varepsilon. \label{eq:coherent-s}
\end{align}
Eq.~\eqref{eq:coherent-s} describes the evolution of the coherent state of the resonator where the average photon occupation is $\bar{n}=|\alpha(t)|^2$ and fluctuates by $\pm \sqrt{\bar{n}}$.
Note that here we implicitly assume the resonator is dressed by the qubit, i.e., $\tilde{\omega}_r$ here refers to the \emph{dressed} resonator frequency and Eq.~\eqref{eq:coherent-s} describes the evolution of the dressed resonator.

As the last step, in order to mimic the behavior of the full JTC ladder at small photon numbers, we modify the interaction term in second line of Eq.~\eqref{eq:H-direct_drive-s} as
\begin{equation}\label{eq:modified_interaction-s}
    \frac{\alpha}{|\alpha|}e^{i(\omega_r - \omega_d)t} \sum_k \text{Re}\left( \sqrt{|\alpha|^2 - k}\right)g_{k, k+1}\ket{k+1}\bra{k} + \text{H.c.}\,.
\end{equation}
In the RWA JTC ladder, only levels that have a fixed number of total excitations can couple to each other, i.e., $|0, \bar{n}\rangle \leftrightarrow |k, \bar{n}-k\rangle$ with total excitation of $\bar{n}$ here.
The above modified drive term prevents interaction between states at the bottom of the JTC ladder where there is not enough total excitations to couple all the levels, i.e., the interaction is turned off when $\bar{n} < k$.
We have checked that with this modification we can exactly reproduce the spectrum of the full qubit-resonator JTC ladder within RWA, specially at small photon numbers.

A reminder that the charge matrix elements $g_{k, k+1}$ inside the transmon cosine well grow roughly as $\sqrt{k}$, and outside of the cosine well they exponentially decrease. 
For simulations, this means if we choose a cut off value for the number of transmon levels that includes a handful of levels outside of the cosine well, we should be safe since the interaction with levels higher than those is exponentially suppressed.

For simplicity and also to match the experimental conditions, in the simulations we use $\omega_d = \omega_r$.
Using a small detuning of a few MHz between resonator and drive, as is the typical case in the dispersive readout, does not change the overall outcome of this model and our results as long as experimental average photon numbers are properly calibrated.
Also note that in simulations we use a square pulse for the resonator drive, i.e., $\varepsilon(t\geq 0) = \varepsilon$, and use a fixed drive amplitude corresponding to a large steady state photon that sweep over all smaller photon numbers as the resonator is populated, i.e., we do not use different drive amplitudes corresponding to different steady state photon numbers.

\begin{figure}[t]
\begin{center}
    \includegraphics[width=0.7\columnwidth]{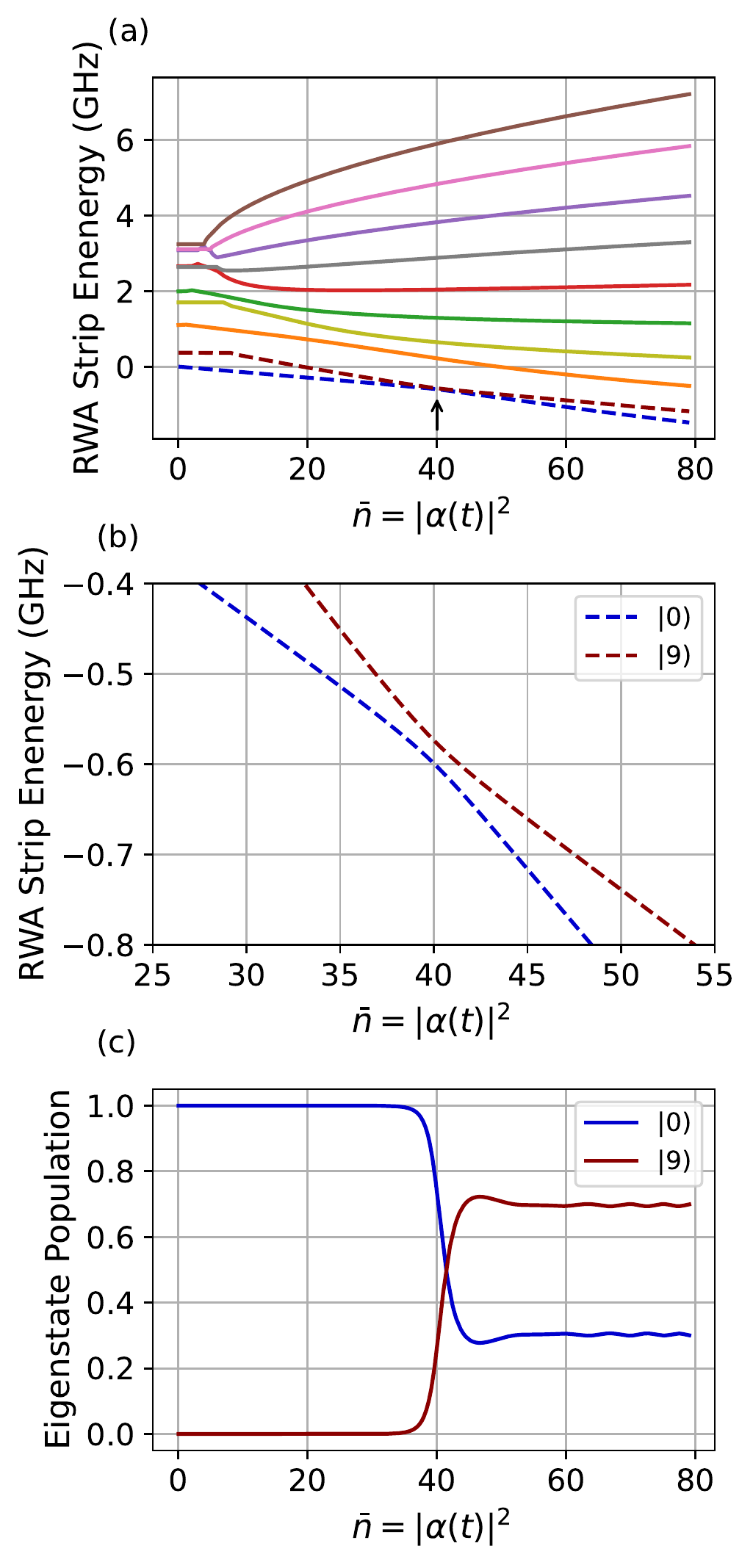}
\end{center}
    \caption{Example of evolution for a directly driven transmon.
    \textbf{(a)} Spectrum of the system as a function of average photon number in the coherent state $\bar{n}=|\alpha(t)|^2$, also called ``fan diagram''.
    Each color belongs to a given transmon level, and not all simulated levels are drawn for brevity.
    Dashed lines show the eigenenergies for $\ket{0}$ (lower dashed line) and $\ket{9}$ (upper dashed line).
    \textbf{(b)} Avoided crossing between $\ket{0}$ and  $\ket{9}$ at $\bar{n} \approx 40$, with splitting of 25 MHz at the crossing.
    \textbf{(c)} Population in the instantaneous eigenstate of the system as a function of average photon number in the coherent state, where the system was initialized in $\ket{0}$.
    A population exchange occurs at the avoided crossing between $\ket{0}$ and $\ket{9}$, which marks a readout transition event.
    All parameters are the same as Fig.~3 of the main paper, while the qubit-resonator detuning and the offset charge are fixed at $\Delta/2\pi=1.1$ GHz and $n_g=0.2$  respectively.
    }
    \label{fig:S3}
\end{figure}

Let us now provide an example of the time evolution of the directly driven transmon, which shows the energy resonances in the system and population exchange between resonant levels.
The procedure for evolving the system via Schr\"{o}dinger equation is outlined in the main text, and here we show an example for a single qubit frequency and offset charge, reiterating that the behavior will be different for other values of these parameters.

Fig.~\ref{fig:S3}(a) shows the spectrum of Eq.~\eqref{eq:H-direct_drive-s} with the modification of  \eqref{eq:modified_interaction-s} as a function of average photon number in the coherent state $\bar{n}$ (time is only a sweeping parameter here such that $\bar{n}=|\alpha(t)|^2$).
This spectrum of the qubit in the rotating frame of the resonator was named ``fan diagram'' in Ref.~\cite{Sank2016}.
Fig.~\ref{fig:S3}(b) shows eigenenergies of $\ket{0}$ and $\ket{9}$, where an avoided crossing occurs between them at around $\bar{n}_\text{cross}=40$, with a splitting of $2g_\text{eff}/2\pi = 25$ MHz at the crossing.
For these parameters, let us use perturbation theory to estimate the effective coupling between these two levels that is mediated via ``virtual'' levels in between them as
\begin{equation}\label{eq:g_eff_purturb-s}
    g_\text{eff} \approx \frac{g_{0,1} g_{1,2} ... g_{8,9}}{\Delta_{1,0} \Delta_{2,0} ... \Delta_{8,0}} \bar{n}_\text{cross} ^ {9/2},
\end{equation}
where $\Delta_{k,0}$ is the detuning between bare levels $k$ and $0$.
One may expect that the detuned levels in between $\ket{0}$ and $\ket{9}$ would suppress the effective coupling between them significantly, however, using the parameters of Fig.~\ref{fig:S3} for Eq.~\eqref{eq:g_eff_purturb-s} we find $g_\text{eff}/2\pi \approx 32$ MHz, which is comparable to the observed value of $12.5$ MHz in Fig.~\ref{fig:S3} and is not suppressed.
Note that as discussed in Refs.~\cite{Khezri2016, Sank2016}, the so called critical photon number $n_\text{crit} \equiv (\Delta/g)^2/4$ \cite{Blais2004} is not the photon number at which the readout transitions occur.
Nevertheless, in order to have large enough effective coupling between resonant levels, we need photon numbers to be larger than $n_\text{crit}$.
This can be seen by noting that the effective coupling of Eq.~\eqref{eq:g_eff_purturb-s} can be rewritten as $g_\text{eff} \propto (\bar{n}_\text{cross}/n_\text{crit})^4 g\sqrt{\bar{n}_\text{cross}}$.

Fig.~\ref{fig:S3}(c) shows the population in the instantaneous eigenstate of the system as a function of average photon number, where the system was initialized in $\ket{0}$.
At the avoided crossing, there is a Landau-Zener type population exchange between $\ket{0}$ and $\ket{9}$, where the loss of population in the initially prepared state $\ket{0}$ signifies a state transition event.
The speed at which the system goes through these avoided crossings is set by the resonator decay rate $\kappa$, meaning an avoided crossing with coupling larger or comparable to $\kappa$ will result in notable population exchange (between diabatic states) and therefore a measurement-induced state transition.

Let us further discuss the effect of resonator energy decay in the full system of Eq.~\eqref{eq:H-lab-s}, where the resonator loses photons at rate $\kappa$, which is usually described via master equation in the Lindblad form \cite{Walls2008}.
This evolution can also be thought as a probabilistic mixture of scenarios described by Kraus operators \cite{Lidar2001}, where a photon loss is described by a stochastic application of annihilation operator $\hat{a}$ onto the system state vector.
Additionally, the system of a transmon coupled to a driven resonator forms a dressed coherent state, i.e., a coherent state made of joint eigenstates of the qubit and resonator.
Therefore, a photon loss event in the full system corresponds to application of the annihilation operator onto a (dressed) coherent state, which by definition retains that state.
This means that the evolution of the full system prior to state transitions (i.e., prior to change of the qubit state) can be described in the eigenbasis of the system with resonator having a coherent state, as we have assumed in this work.
Another perspective on the same problem that validates our semi-classical approximation is to note that a photon loss moves the system from one RWA strip to another, and with large enough photons in the system these RWA strips are approximately similar copies of each other and yield the same evolution for the qubit.


%

\end{document}